# Influence of the density of states of graphene on the transport properties of graphene/MoS$_2$/metal vertical field-effect transistors

Rai Moriya[1,a)], Takehiro Yamaguchi[1], Yoshihisa Inoue[1], Yohta Sata[1], Sei Morikawa[1], Satoru Masubuchi[1,2], and Tomoki Machida[1,2,b)]

[1]*Institute of Industrial Science, University of Tokyo, 4-6-1 Komaba, Meguro, Tokyo 153-8505, Japan*
[2]*Institute for Nano Quantum Information Electronics, University of Tokyo, 4-6-1 Komaba, Meguro, Tokyo 153-8505, Japan*

We performed detailed studies of the current–voltage (*I–V*) characteristics in graphene/MoS$_2$/metal vertical field-effect transistors. Owing to its low density of states, the Fermi level in graphene is very sensitive to its carrier density and thus the external electric field. Under the application of a bias voltage $V_B$ between graphene and the metal layer in the graphene/MoS$_2$/metal heterostructure for driving current through the van der Waals interface, the electric field across the MoS$_2$ dielectric induces a shift in the Fermi level of graphene. When the Fermi level of graphene coincides with the Dirac point, a significant nonlinearity appears in the measured *I–V* curve, thus enabling us to perform spectroscopy of the Dirac point. By detecting the Dirac point for different back-gate voltages, we revealed that the capacitance of the nanometer-thick MoS$_2$ layer can be determined from a simple DC transport measurement.

a)E-mail: moriyar@iis.u-tokyo.ac.jp
b)E-mail: tmachida@iis.u-tokyo.ac.jp

Owing to recent advances in fabrication techniques, van der Waals (vdW) heterostructures of graphene and other two-dimensional crystals such as hexagonal boron nitride (h-BN) and transition-metal dichalcogenides (TMDs) have become possible [1]. Among these materials, graphene has a unique Dirac-like band structure; this enables continuous tuning of its carrier concentration from electrons to holes [2]. Moreover, the Fermi level of graphene significantly changes with the carrier concentration owing to its low density of states [3]. This gate tunability of the Fermi energy of graphene has been used for an electrode in the vdW heterostructure of graphene and other two-dimensional materials [4-11]. Vertical field-effect transistors of graphene/h-BN/graphene [4,11], graphene/WS$_2$/graphene [5,6], and graphene/MoS$_2$/metal [7,8,12] have been already demonstrated and received considerable attention for electronics applications. In such vertical heterostructure devices, the application of a source–drain bias to graphene and another electrode (either graphene or metal) could induce a change in the carrier concentration in the graphene electrode due to the capacitive coupling across the dielectric h-BN or TMD layer. Thus, even for a constant back-gate voltage $V_G$, the Fermi level in graphene continuously changes during a current–voltage ($I$–$V$) measurement. Because there is a singularity in the density of states at the Dirac point, the current flow across the junction is expected to change when the Fermi level coincides with the Dirac point. The contribution of the Dirac-point singularity to the conductance has been studied for a graphene/h-BN/graphene tunnel transistor [4,11]; in these devices, the tunnel conductance itself is strongly influenced by the density of states of graphene. However, the contribution of the Dirac-point singularity to the conductance has not been studied for graphene/TMD vdW heterostructures. In these devices, the transport is dominated by thermionic emission across the vdW interface, and experimentally, a larger ON–OFF current ratio and a large ON current density have been demonstrated [5,7,12]; thus, they are more suitable from an application viewpoint. Here, we studied the influence of the density of states of graphene on the transport properties of a graphene/MoS$_2$/metal vertical field-effect transistor and found that the energy of the Dirac point can be detected from simple two-terminal $I$–$V$ curve measurements.

The device structure is schematically illustrated in Fig. 1(a). First, monolayer graphene is fabricated by mechanical exfoliation from Kish graphite and deposited onto a 300-nm-thick SiO$_2$/$n^+$-Si substrate. Then, the MoS$_2$ layer is exfoliated from a bulk MoS$_2$ crystal (2D Semiconductors Inc.) and deposited onto the graphene layer using the dry transfer method [12]. The freshly cleaved surfaces of graphene and the MoS$_2$ layer are contacted through the vdW force. Subsequently, metal electrodes (30 nm Au/50 nm Ti) are fabricated using standard electron beam (EB) lithography (Elionix ELS7500) and EB evaporation. The $I$–$V$ characteristics of the device are measured by applying a source–drain bias voltage $V_B$ between the metal and graphene, and the current flow $I$ across the vdW junction is measured. The number of MoS$_2$ layers, $N$, is calculated by assuming that the monolayer thickness of MoS$_2$ is 0.65 nm. The junction areas for the series of fabricated devices are in the range of 1–3 μm$^2$.

The band structures of the graphene/MoS$_2$/metal heterostructure are illustrated in Fig. 1(b). The Fermi level of the graphene layer can be adjusted with the application of a back-gate voltage $V_G$. The band alignment for $V_G < 0$ and $V_B = 0$ V is illustrated in the left panel of Fig. 1(b). The application of a $V_B$ drives the current flow vertically into the vdW heterostructure; previous studies revealed that thermionic-emission (TE) is the dominant conduction mechanism in this device [7,12,13]. The application of $V_B$ changes the Fermi level in graphene because of capacitive coupling through $MoS_2$, as depicted in the middle panel of Fig. 1(b). By increasing $V_B$ further, the Fermi level of graphene continuously shifts, and the Fermi level of graphene coincides with the Dirac point at some point, as shown in the right panel of Fig. 1(b). The conduction across the vertical heterostructure changes owing to the singularity in the density of states at the Dirac point (this mechanism will be discussed more detail later). Therefore, the position of the Dirac point can be probed by a simple two-terminal conductance measurement. Interestingly, this coincidence condition corresponds to the fact that the carriers in graphene induced by the back-gate voltage and background impurities are exactly canceled by the carriers induced by the source–drain voltage. By measuring this coincidence condition for different values of $V_G$, the ratio of the back-gate capacitance to that of $MoS_2$ can be obtained. As a result, the dielectric constant of $MoS_2$ can be determined from the $V_G$ dependence of the $I$–$V$ curve measurement.

An atomic force microscopy (AFM) topography image of a typical device is shown in Fig. 2(a). The number of $MoS_2$ layers for this device is $N = 37$. The Au/Ti metal contacts are fabricated on both ends of graphene strip. By using these contacts, the two-terminal resistance of graphene partially covered with a $MoS_2$ film as a function of $V_G$ is shown in Fig. 2(b). The Dirac point located at $V_G \sim +5$ V is clearly visible. Next, the vertical transport across graphene/$MoS_2$/Ti is measured as a function of $V_G$, and the results are shown in Fig. 2(c). The current is normalized to the junction area of 1 $\mu m^2$. The $I$–$V$ curves are strongly modified with $V_G$; this is due to the modulation of the Schottky barrier at the graphene/$MoS_2$ interface, as reported previously [7,12]. At $V_G = -50$ V, the Fermi level of graphene is the lowest; thus, the Schottky barrier height is the highest, and the conductance across the graphene/MoS$_2$ interface is suppressed. With increasing $V_G$, the Fermi level of the graphene layer increases; as a result, the Schottky barrier height at the graphene/MoS$_2$ interface decreases, and the conductance increases. These results reveal that the basic transport property of the graphene/MoS$_2$/Ti vdW heterostructure is a gate-controlled Schottky barrier. With this gate-tunable conductance, we noticed an inflection point in the $I$–$V$ curve, as indicated by the arrow; interestingly, its position also changes with $V_G$. A contour plot of the differential conductance $d(\log I)/dV_B$ for sweeps of both $V_B$ and $V_G$ is shown in Fig. 2(d). The inflection point linearly changes, as indicated by the dashed line, and this line crosses the point at $V_G = +6$ V and $V_B = 0$ V, which is close to the Dirac point of the bottom graphene electrode [Fig. 2(b)]. The slope of this line is $V_G/V_B = -24$.

A numerical calculation based on the electrostatics of the vertical heterostructure is performed. The device can be regarded as metal plates with dielectrics between the plates, as illustrated in Fig. 2(e). In this figure, $\varepsilon_{SiO_2}$, $\varepsilon_{MoS_2}$, $d_{SiO_2}$, and $d_{MoS_2}$ denote the dielectric constants and thicknesses of SiO$_2$ and MoS$_2$. $F_G$ and $F_1$ denote the electric field on the back-gate and MoS$_2$ sides of the graphene layer, respectively. $n_G$ and $n_M$ are the carrier concentrations of the graphene and Ti layers, respectively. The following series of equations are obtained from the Fig. 2(e) [4,7]:

$$\begin{cases} \varepsilon_{SiO_2}F_G - \varepsilon_{MoS_2}F_1 = -n_G e \\ \varepsilon_{MoS_2}F_1 = -n_M e \\ eV_B = e\Delta V - \mu_G\left(n_G\right) + W_{GM} \end{cases}, \qquad (1)$$

where $F_G = V_G / d_{SiO_2}$, $F_1 = \Delta V / d_{MoS_2}$, $W_{GM}$ denotes the work-function difference between graphene at charge neutrality and Ti, and $e$ is the elementary charge. The relationship between $V_B$, $\Delta V$, and $\mu_G$ is illustrated in Fig. 1(b). The Fermi level of graphene is given by $\mu_G(n_G) = \text{sign}(n_G)\hbar v_F\sqrt{\pi|n_G|}$, where $\hbar$ denotes the Planck constant and $v_F = 1 \times 10^6$ m/s the graphene's Fermi velocity. From these equations, we are able to calculate $\mu_G(n_G)$ for given values of $V_G$ and $V_B$. The transport across the graphene/MoS$_2$ interface can be calculated with the thermionic-emission model [14]:

$$J = A^* T^2 \exp\left(-\frac{e\varphi_B}{k_B T}\right)\left[-\exp\left\{\frac{e(V_B - R_s J)}{k_B T}\right\} + 1\right], \tag{2}$$

where $A^* = em^* k_B^2/(2\pi^2\hbar^3)$ represents the effective Richardson constant, $m^*$ is the effective mass of an electron in MoS$_2$, $k_B$ is Boltzmann's constant, $R_s$ is the series resistance of the device, $\varphi_B = \varphi_0 - \mu_G - \varphi_{IF}$ is the effective Schottky barrier height, $\varphi_0$ is the band offset between MoS$_2$ and graphene at charge neutrality, $\varphi_{IF} = \sqrt{eE_{in}/4\pi\varepsilon_{MoS_2}}$ is the image-force lowering of the Schottky barrier height, $E_{in} = (\varphi_0 - \mu_G - \varphi_M)/d$ is the internal electric field in the MoS$_2$ layer, and $\varphi_M$ is the Schottky barrier height between MoS$_2$ and Ti. From the lateral transport in the MoS$_2$ layer, we determined $\varphi_M = 0.13$ eV. We assumed that the MoS$_2$ layer is fully depleted in the thickness range we studied [7,8,12], and we chose $\varphi_0 = 0.35$ eV, $W_{GM} = 0.2$ eV, and $m^* = 0.6m_0$ [15]. The calculated $I$–$V$ curve for $N = 37$ and $\varepsilon_{MoS_2} = 7.4$ is presented in Fig. 2(f). Compared with Fig. 2(c), the calculated curve exhibits good agreement, and noticeably, the inflection point in the $I$–$V$ curve is also reproduced in our calculation, indicating the validity of our model. A contour plot of the differential conductance $d(\log I)/dV_B$ is plotted in Fig. 2(g). The calculation based on simple electrostatics exhibits good agreement with the experimental results. The inflection point corresponds to when the Fermi level of graphene is aligned with the Dirac point, and $d\mu_G/dV_B$ has its maximum value at this point. Note that our calculation did not take into account the direct contribution of the density of states of graphene to the conductivity of the vertical heterostructure, as done for the graphene-based tunneling device [4,11]. Nevertheless, the calculated results exhibit good agreement with the experimental results, suggesting that electrostatic coupling across the dielectrics is the dominant contribution to the transport for thermionic-emission across the graphene/MoS$_2$ Schottky barrier. The linear slope of the inflection point in the $V_G$−$V_B$ mapping in Figs. 2(d) and (g) can be obtained by setting $\mu_G = 0$ and solving Eq. (1). We obtained following expression:

$$V_G = \left( \frac{d_{SiO_2} \varepsilon_{MoS_2}}{d_{MoS_2} \varepsilon_{SiO_2}} \right) V_B + \frac{d_{SiO_2} \varepsilon_{MoS_2} W_{GM}}{e d_{MoS_2} \varepsilon_{SiO_2}}. \tag{3}$$

Equation (3) indicates that the relationship between $V_G$ and $V_B$ is linear. The slope is given by the ratio between the geometrical capacitance of SiO$_2$, $C_{SiO_2} = \varepsilon_{SiO_2}/d_{SiO_2}$ , and the geometrical capacitance of MoS$_2$, $C_{MoS_2} = \varepsilon_{MoS_2}/d_{MoS_2}$ . We determined the ratio $C_{MoS_2}/C_{SiO_2} = 24$ from Fig. 2(d). Note that the linear relationship obtained in Eq. (3) is due to the fact that metal is used as one of the electrodes. Otherwise, if graphene is used for both electrodes, the solution of Eq. (1) becomes highly nonlinear because of the strong quantum capacitance contribution [11]. With the metal electrode, the position of the Dirac point can be related to the geometrical capacitance only, thereby resulting in the linear relationship expressed in Eq. (3).

We fabricated a series of graphene/MoS$_2$/metal vertical heterostructures with different MoS$_2$ thicknesses and measured the $V_G$ dependence of the $I$–$V$ characteristics. Contour plots of $d(\log I)/dV_B$ for voltage sweeps of both $V_B$ and $V_G$ for $N = 37$, 21, and 13 are shown in Fig. 3(a). We observed a linear change in the inflection point with respect to $V_G$ and $V_B$ for all devices. Noticeably, the slope of the line changes with $N$; we observed a larger value for $V_G/V_B$ for a thinner MoS$_2$ layer. This suggests $C_{MoS_2}/C_{SiO_2}$ increases for thinner MoS$_2$ layers. The relationship between $C_{MoS_2}/C_{SiO_2}$ and $N$ is summarized in Fig. 3(b). Because $C_{SiO_2}$ is constant among all devices, the results in Fig. 3(b) revealed a change in $C_{MoS_2}$ with $N$. Assuming the bulk value of $\varepsilon_{MoS_2} = 8.5$ [16], the relationship between $C_{MoS_2}/C_{SiO_2}$ and $N$ is calculated and plotted in Fig. 3(b). Although $C_{MoS_2}/C_{SiO_2}$ exhibits good agreement with bulk value for thicker MoS$_2$ layers, it deviates for thinner MoS$_2$ layers. By using $C_{SiO_2} = 113.6$ μF/m$^2$ for 300-nm-thick SiO$_2$, we calculated $\varepsilon_{MoS_2}$ with respect to $N$, and the results are shown in Fig. 3(c). The results reveal that $\varepsilon_{MoS_2}$ exhibits a value close to the bulk value for thick MoS$_2$ layers; however, it decreases for thinner MoS$_2$ layers. According to theoretical calculations performed by other groups [16,17], $\varepsilon_{MoS_2}$ is expected to decrease for lower values of $N$ owing to the quantum confinement effect. We think our experimental results indeed exhibit such a layer dependence for $\varepsilon_{MoS_2}$. Similar $N$ dependence of $\varepsilon_{MoS_2}$ is reported by using the conventional capacitance measurement on MoS$_2$ films [18].

In summary, we revealed the influence of the low density of states of graphene on the transport in a graphene/MoS$_2$/metal vertical heterostructure. The Dirac point of the graphene layer can be probed from a simple DC $I$–$V$ measurement owing to the singularity in the density of states for graphene. The use of a metal electrode on one side of the structure allows for the determination of the capacitance of 5-to-42-monolayer-thick $MoS_2$, and therefore, its dielectric constant. The results and analysis presented here do not depend on the type of dielectric material. Our findings reveal a way of utilizing the low density of states of a graphene electrode for nanoelectronics applications.

**Acknowledgements**

This work was partly supported by a Grant-in-Aid for Scientific Research on Innovative Areas "Science of Atomic Layers" of the Ministry of Education, Culture, Sports, Science and Technology (MEXT) and the Project for Developing Innovation Systems of MEXT and Grants-in-Aid for Scientific Research from the Japan Society for the Promotion of Science (JSPS). S. Morikawa acknowledges the JSPS Research Fellowship for Young Scientists.

**Figure captions**

Figure 1:

(a) Schematic of a graphene/MoS$_2$/metal vertical heterostructure and the measurement circuit. (b) Band alignment of the vertical heterostructure device. The left panel represents the application of $V_G < 0$, the middle panel represents the application of $V_B$, and the right panel represents a further increase in $V_B$.

Figure 2:

(a) AFM topography image of a graphene/MoS$_2$/metal heterostructure when the number of MoS$_2$ layers $N = 37$. (b) Two-terminal resistance of the graphene layer as a function of $V_G$ measured between contact A and B in Fig. 2(a). (c) $I$–$V$ curves of vertical transport in a graphene/MoS$_2$/metal heterostructure measured at different values of $V_G$. (d) Differential conductance $d(\log I)/dV_B$ as a function of $V_G$ and $V_B$ calculated from the $I$–$V$ curves in Fig. 2(c). (e) Electrostatics of the heterostructure. MG denotes the back-gate electrode, Gr the graphene, and M the Au/Ti contact on the MoS$_2$. (f) Calculated $I$–$V$ curves for the graphene/MoS$_2$/metal heterostructure. (g) Calculated $d(\log I)/dV_B$ as a function of $V_G$ and $V_B$.

Figure 3:

(a) Differential conductance $d(\log I)/dV_B$ as a function of $V_G$ and $V_B$ for different numbers of MoS$_2$ layers $N = 37$, 21, and 13. (b) Ratio between the capacitances of MoS$_2$ and SiO$_2$, $C_{\mathrm{MoS_2}}/C_{\mathrm{SiO_2}}$, as a function of the $\mathrm{MoS_2}$ layer thickness. The dashed line represents the calculated $C_{\mathrm{MoS_2}}/C_{\mathrm{SiO_2}}$ for $\varepsilon_{\mathrm{MoS_2}} = 8.5$. (c) $\varepsilon_{\mathrm{MoS_2}}$ versus the number of $\mathrm{MoS_2}$ layers, $N$.

Figure 1

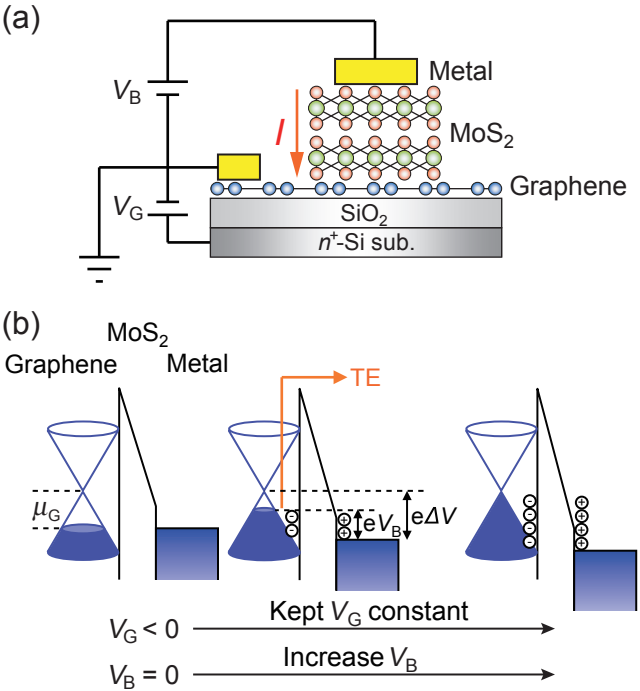

(a)

$V_B$

$V_G$

Metal

MoS$_2$

Graphene

SiO$_2$

$n^+$-Si sub.

$I$

(b)

Graphene    MoS$_2$    Metal

TE

$\mu_G$

$eV_B$    $e\Delta V$

$V_G < 0$ ——————→ Kept $V_G$ constant

$V_B = 0$ ——————→ Increase $V_B$

Figure 2

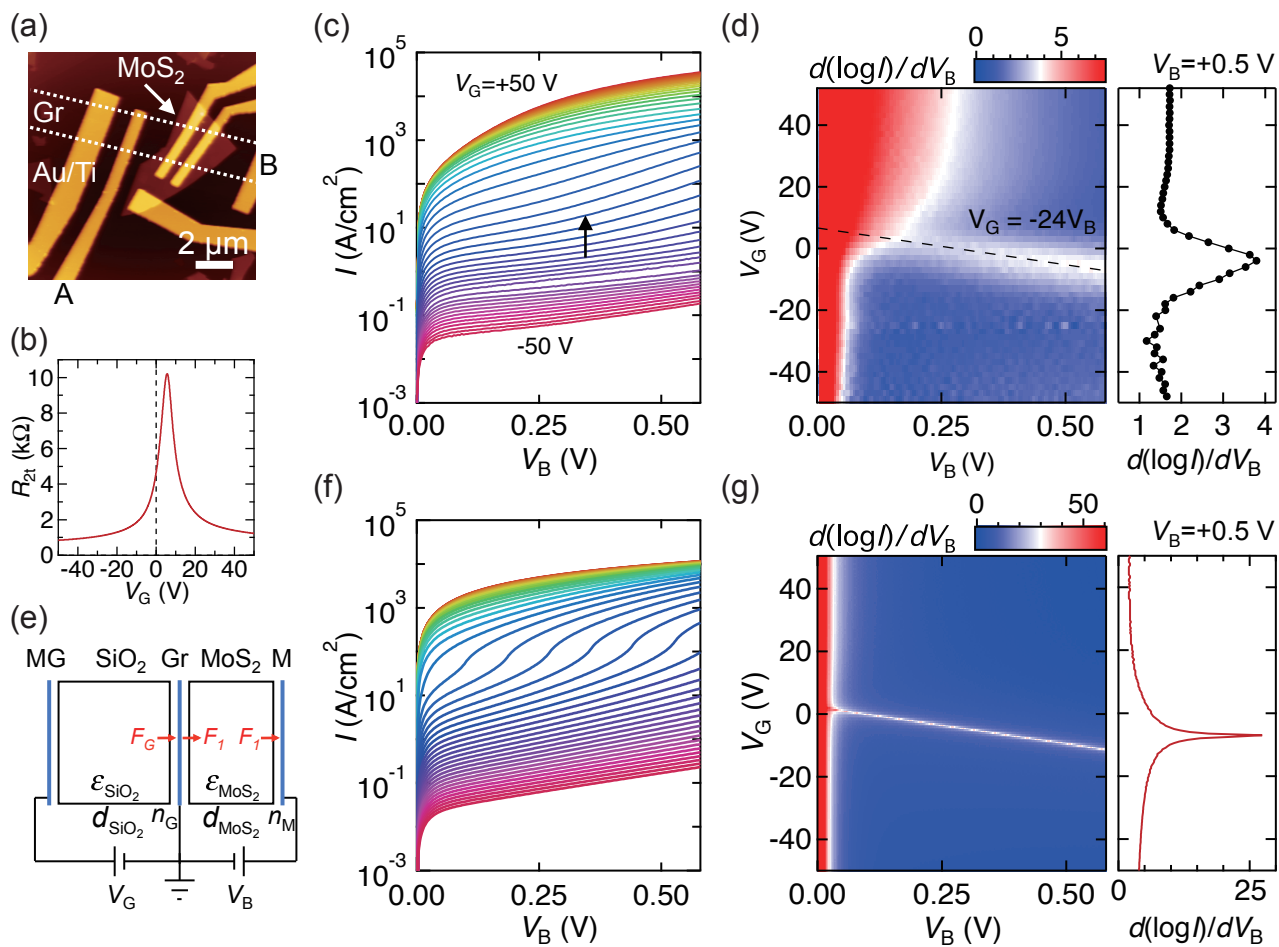

Figure 3

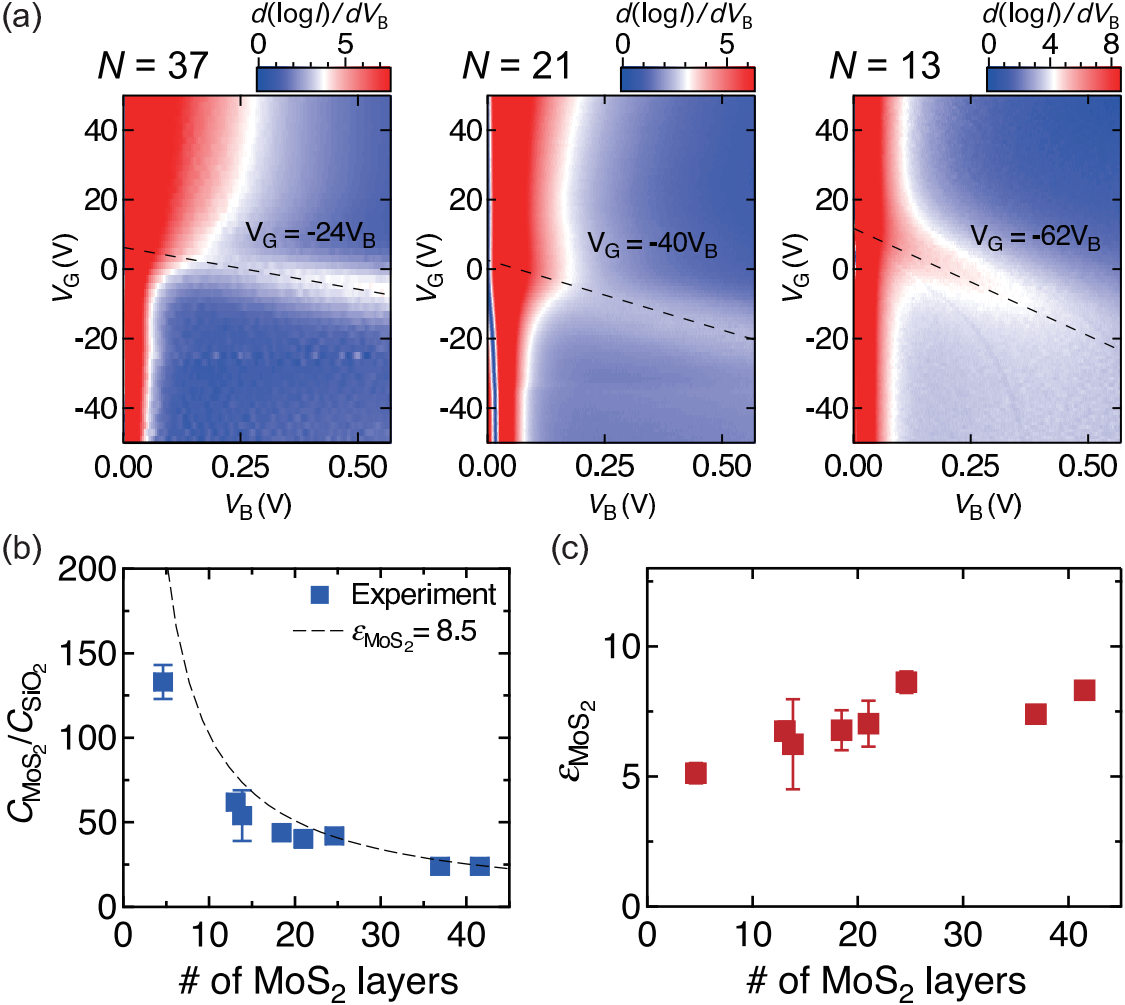